\documentclass[aps,prl,twocolumn,showpacs,superscriptaddress,floatfix,nofootinbib]{revtex4-1}

\usepackage{graphicx}
\usepackage{amsmath}
\usepackage{epsfig}
\usepackage{helvet}
\usepackage{amssymb}

\newcommand{\be}{\begin{equation}}
\newcommand{\ee}{\end{equation}}
\newcommand{\bea}{\begin{eqnarray}}
\newcommand{\eea}{\end{eqnarray}}

\newcommand{\tr}{\mbox{tr}}
\newcommand{\bra}[1]{\mbox{$\langle #1 |$}}
\newcommand{\ket}[1]{\mbox{$| #1 \rangle$}}

\def\tr{ \mbox{tr}}

\def\Wdec{V}
\def\Wcoa{W}

\begin{document}

\title{A real space decoupling transformation for quantum many-body systems}
\author{G. Evenbly}
\affiliation{Institute for Quantum Information, California Institute of Technology, MC 305-16, Pasadena CA 91125, USA}
\author{G. Vidal}
\affiliation{Perimeter Institute for Theoretical Physics, Waterloo, Ontario, N2L 2Y5, Canada}
\date{\today}

\begin{abstract}
We propose a real space renormalization group method to explicitly decouple into independent components a many-body system that, as in the phenomenon of spin-charge separation, exhibits separation of degrees of freedom at low energies. Our approach produces a branching holographic description of such systems that opens the path to the efficient simulation of the most entangled phases of quantum matter, such as those whose ground state violates a boundary law for entanglement entropy.
As in the coarse-graining transformation of [Phys. Rev. Lett. 99, 220405 (2007)], the key ingredient of this decoupling transformation is the concept of entanglement renormalization, or removal of short-range entanglement.
We demonstrate the feasibility of the approach, both analytically and numerically, by decoupling in real space the ground state of a critical quantum spin chain into two. Generalized notions of RG flow and of scale invariance are also put forward.
\end{abstract}

\maketitle

At low energies, an extended quantum many-body system may decompose into several independent components, as in the phenomenon of spin-charge separation, whereby a system of interacting electrons decouples into spinons and holons \cite{SpinCharge}.
Another example is a Fermi liquid in $D\geq 2$ dimensions, whose Fermi surface factorizes into an infinite collection of one-dimensional critical systems \cite{FermiSurface}. More exotic systems, such as spin Bose-metals displaying a Bose surface \cite{BoseMetal}, are also believed to decouple in a similar way. In the above examples, the factorization of the many-body Hilbert space into
components at low energies becomes evident when the Hamiltonian is expressed in terms of the adequate \textit{collective} quasi-particle operators in \textit{momentum space}. However, a generic interacting many-body system is naturally modeled by some \textit{microscopic} degrees of freedom with a local Hamiltonian in \textit{real space}, where this factorization may no longer be manifest.

The purpose of this paper
is to propose a theoretical framework and a numerical technique to obtain, in real space, a factorized description of the ground state (more generally, low energy subspace) of a many-body system that experiences separation of degrees of freedom at low energies. There are several good reasons, both conceptual and computational, to search for one such description. Fermi liquids and spin-Bose metals, which are among the most entangled phases of quantum matter, pose an exceptional challenge to real space renormalization group (RG) methods. Recall that in $D\geq 2$ dimensions, where the ground states of many phases (both gapped and gapless) conform to a boundary law for entanglement entropy \cite{BoundaryLaw}, the ground states of systems with emergent Fermi or Bose surfaces display logarithmic violations of this boundary law \cite{Violation}. A logarithmic correction may seem small business. However, it has the dramatic effect, as one flows towards large length scales, of increasing the number of degrees of freedom that one needs to deal with, thus rendering real space RG methods inefficient. In particular, it precludes the use of all known forms of tensor network ans\"atze \cite{PEPS,MERA}, which can only efficiently describe ground states that strictly obey an entropic boundary law.
This is most unfortunate, since tensor networks have recently emerged both as the basis of powerful non-pertubative approaches to many-body systems on the lattice \cite{PEPS,MERA,CFT,newPEPS} and as the natural language to describe and classify exotic phases of quantum matter \cite{Classify}.
An ultimate goal of our work is to understand the entanglement pattern in systems, such as Fermi Liquids and spin-Bose metals, that violate the entropic boundary law, and to extend the tensor network formalism to these exceptionally entangled phases of quantum matter.

Specifically, here we will show that entanglement renormalization, or removal of short-range entanglement ---already the basis of a renormalization group transformation for quantum systems on a lattice \cite{ER}---, can also be used to identify and decouple, in real space, different components of a system that experiences separation of degrees of freedom. For illustrative purposes, we will demonstrate the performance of the approach in $D=1$ dimensions only, by numerically decoupling a critical quantum spin chain into two independent components. However, most of our discussion also readily applies to $D\geq 2$ dimensions. Decoupling in real space leads to a generalized multi-scale entanglement renormalization ansatz (MERA), the \textit{branching} MERA \cite{BranchingMERA}, which is a tensor network ansatz capable of reproducing violations of the entropic boundary law. It also defines a holographic geometry \cite{Geometry} with branches for each of the components, drastically reducing computational costs, and suggesting a generalized notion of scale invariance.

Let $\mathcal{L}$ be a one-dimensional lattice made of $N$ sites, where each site is described by a Hilbert space $\mathbb{V}$ of finite dimension $\chi$. We denote by $H$ and $\ket{\Psi}$ a local Hamiltonian acting on $\mathbb{V}^{\otimes N}$ and its ground state, respectively. Here we are interested in a very particular class of lattice models. Namely, we will assume that the model can be decoupled into two models, with lattices $\mathcal{L}_A$ and $\mathcal{L}_B$, each made of $N/2$ sites, under a suitable \textit{decoupling} transformation $\Wdec$ to be introduced below.

\begin{figure}[!tb]
\begin{center}
\includegraphics[width=7cm]{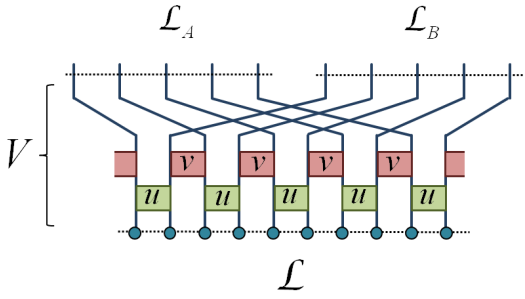}
\caption{
(color online) Decoupling transformation $V$ for a one-dimensional lattice $\mathcal{L}$. $V$ maps the lattice model into two decoupled lattice models, with lattices $\mathcal{L}_A$ and $\mathcal{L}_B$. The decoupling transformation $V$ decomposes as the product of two rows of (two-site) disentanglers $u$ and $v$, and a number of (two-site) swap gates that bring all the sites of sublattice $\mathcal{L}_A$  ($\mathcal{L}_B$) together to the left (respectively, right).
} \label{fig:decoupling}
\end{center}
\end{figure}

Let us first clarify what we mean by 'decoupling' a lattice model into two, since this term may be used to describe several inequivalent scenarios. The strongest notion refers to 'Hamiltonian decoupling': $\Wdec$ maps the Hamiltonian $H$ into a sum of two Hamiltonians $H_A$ and $H_B$ for sublattices $\mathcal{L}_A$ and $\mathcal{L}_B$,
\begin{equation}
	H \rightarrow H' \equiv \Wdec^{\dagger} H \Wdec = H_A + H_B.
\label{eq:fact_H}
\end{equation}
Notice that the ground state $\ket{\Psi}$ is then mapped into the product of ground states for $H_A$ and $H_B$,
\begin{equation}
 \ket{\Psi} \rightarrow	\ket{\Psi'} \equiv \Wdec^{\dagger}\ket{\Psi} = \ket{\Psi_A}\otimes \ket{\Psi_B}.
\label{eq:fact_psi}
\end{equation}
A second, weaker notion of decoupling is that of 'ground state decoupling', where the transformed Hamiltonian $H'$ may contain interactions between sublattices but, nevertheless, its ground state $\ket{\Psi'}$ still factorizes as in Eq. \ref{eq:fact_psi}.
We note for later reference that in this case $\ket{\Psi_A}$ and $\ket{\Psi_B}$ are also the ground states of the reduced Hamiltonians $H'_A$ and $H'_B$,
\begin{equation}\label{eq:fact_H2}
    H'_A \equiv \bra{\Psi_B} H' \ket{\Psi_B},~~~~~~ H'_B \equiv \bra{\Psi_A} H' \ket{\Psi_A}.
\end{equation}

\begin{figure}[!tb]
\begin{center}
\includegraphics[width=8.5cm]{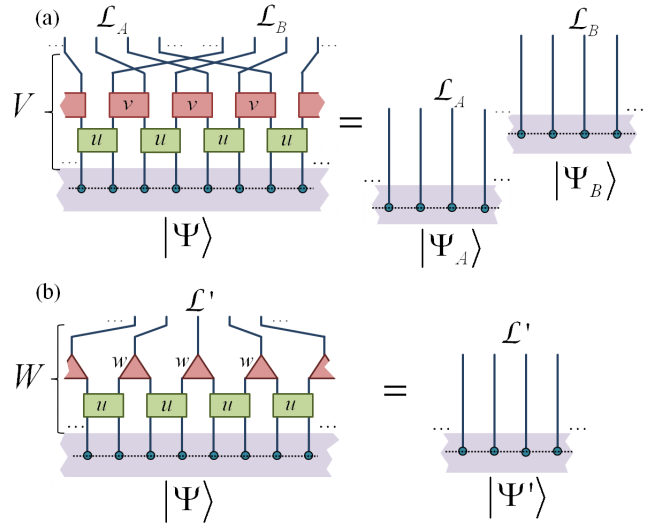}
\caption{
(color online) (a) Decoupling transformation $V$, mapping the ground state $\ket{\Psi}$ of a local Hamiltonian $H$ on lattice $\mathcal{L}$, into the tensor product of states $\ket{\Psi_A}$ and $\ket{\Psi_B}$ for lattices $\mathcal{L}_A$ and $\mathcal{L}_B$, as in Eq. \ref{eq:fact_psi}. (b) Coarse-graining transformation $W$, mapping the ground state $\ket{\Psi}$ of a local Hamiltonian $H$ on lattice $\mathcal{L}$ into the state $\ket{\Psi'}$ of a coarse-grained lattice $\mathcal{L}'$.
} \label{fig:VandW}
\end{center}
\end{figure}

\textit{Decoupling transformation}.--- Our decoupling transformation $\Wdec$ is a unitary transformation implemented in real space by means of local gates, as depicted in Fig. \ref{fig:decoupling}. It is made of two types of local gates: (i) \textit{disentanglers}, that in the present example appear organized in two rows; and (ii) \textit{swaps}, used to regroup the sites of sublattices $\mathcal{L}_A$ and $\mathcal{L}_B$. Disentanglers $u$ and $v$ are two-site transformations, i.e. $u:\mathbb{V}^{\otimes 2} \rightarrow \mathbb{V}^{\otimes 2}$ and $v:\mathbb{V}^{\otimes 2} \rightarrow \mathbb{V}^{\otimes 2}$, constrained to be unitary, i.e. $uu^{\dagger} = u^{\dagger}u = vv^{\dagger} = v^{\dagger}v = \mathbb{I}^{\otimes 2}$,
with $O(\chi^4)$ complex coefficients, which can be treated as variational parameters (to be determined, for instance, through numerically optimization). The role of disentanglers is to remove short-range entanglement between sublattices $\mathcal{L}_A$ and $\mathcal{L}_B$ so as to produce a factorized ground state, Eq. \ref{eq:fact_psi} and Fig. \ref{fig:VandW}(a), or to eliminate interaction terms that couple the two sublattices to produce a decoupled Hamiltonian, Eq. \ref{eq:fact_H}. Disentanglers are then followed by swap gates, each of which permutes the basis states of a pair of lattice sites. For bosonic degrees of freedom a swap gate acts just as
\begin{equation}
	\mbox{swap}:\ket{i}\otimes\ket{j} \rightarrow \ket{j}\otimes\ket{i}, ~~~~~~~~~~~~~~ (\mbox{bosons})
\label{eq:swap_b}
\end{equation}
where $\{\ket{i}\}$ denotes a basis in $\mathbb{V}$, whereas for fermionic degrees of freedom it includes a conditional minus sign,
\begin{equation}
\mbox{swap}:\ket{i}\otimes\ket{j} \rightarrow (-1)^{f(i,j)}\ket{j}\otimes\ket{i}, ~~ (\mbox{fermions})
\label{eq:swap_f}
\end{equation}
where $f(i,j)=1$ if both $\ket{i}$ and $\ket{j}$ carry an odd fermionic particle number and $f(i,j)=0$ otherwise. For anyonic degrees of freedom, the swap gates will embody the appropriate representation of the braid group.

A key question is what type of many-body systems are susceptible to decouple, in the sense of Eqs. \ref{eq:fact_H} or \ref{eq:fact_psi}, under the decoupling transformation $\Wdec$. A concrete example is given by the XX quantum spin chain,
\begin{equation}
	H^{\tiny \mbox{XX}} = \sum_r \left(X_rX_{r+1} + Y_rY_{r+1} + \mu/2 Z_r\right),
	\label{eq:XX}
\end{equation}
where $X,Y$ and $Z$ are Pauli matrices. As discussed in Ref. \cite{Yellow} (exercise 12.1, pp 480), for $\mu=0$ this Hamiltonian can be \textit{exactly} decoupled into two independent copies of the quantum Ising chain with critical transverse magnetic field,
\begin{equation}
	H^{\tiny \mbox{Ising}} = \sum_r \left(X_rX_{r+1} +  Z_r\right).
	\label{eq:Ising}
\end{equation}
Moreover, as can be easily seen by rewriting $H^{\tiny \mbox{XX}}$ in Majorama fermionic modes \cite{CriticalChain}, this Hamiltonian can be decoupled as in Eq. \ref{eq:fact_H} by a transformation $V$ made of analytical disentanglers (that simply permute some of these modes) and the fermionic swaps of Eq. \ref{eq:swap_f} (see Appendix).

\begin{figure}[!tb]
\begin{center}
\includegraphics[width=8.5cm]{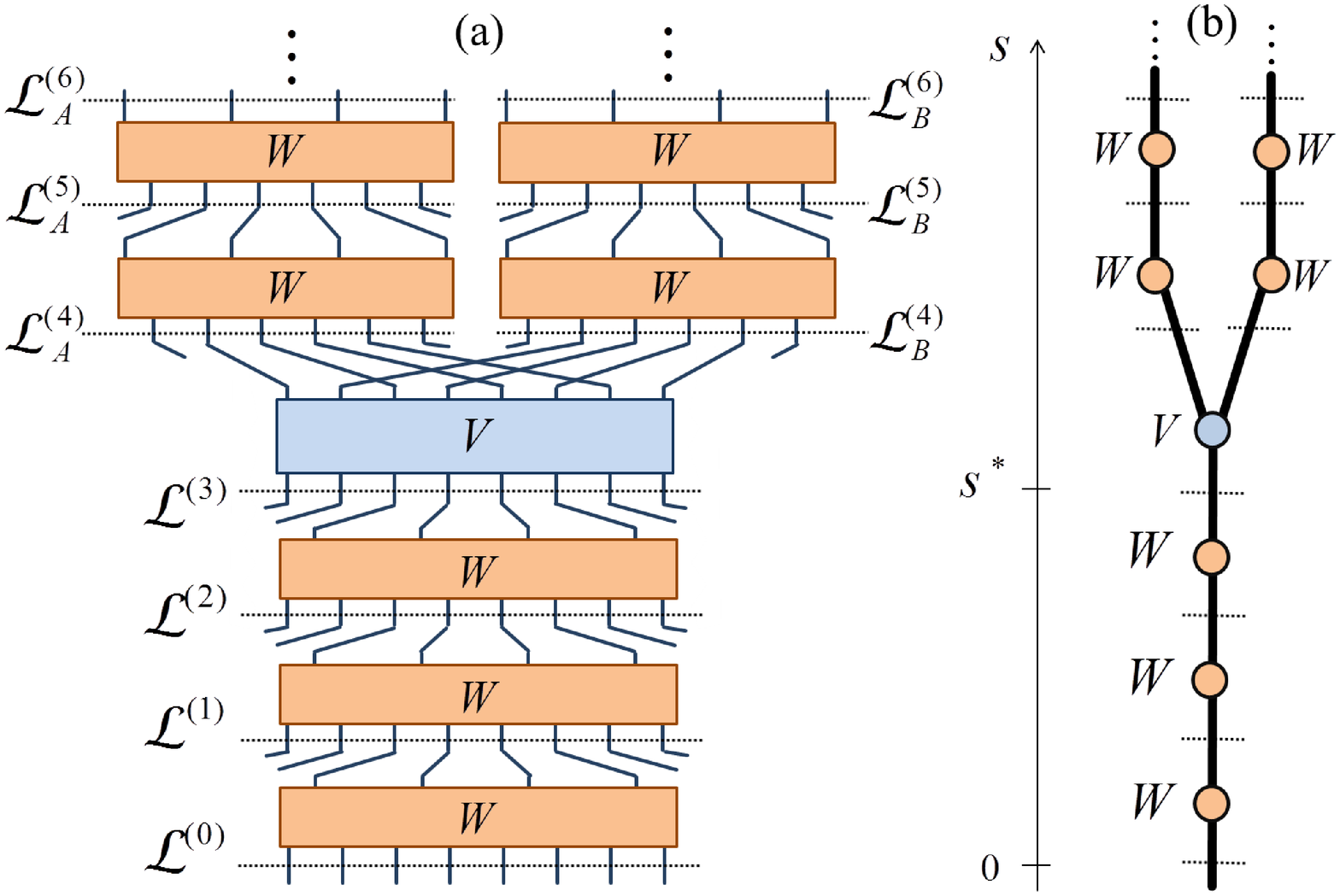}
\caption{
(color online) (a) Sequence of coarse-graining transformations $W$ and decoupling transformation $V$ for a system with separation of degrees of freedom, with $s^{\star}=3$. Each transformation $\Wdec$ or $\Wcoa$ decomposes into two-site gates as in Fig. \ref{fig:VandW}. The resulting quantum circuit defines the branching MERA. (b) Corresponding holographic tree. The decoupling transformation $V$ produces branching in the holographic or scale direction (vertical axis).
} \label{fig:Y}
\end{center}
\end{figure}

\textit{Holographic branching.---} Generically, however, separation of degrees of freedom is likely to take place at a length scale, the \textit{decoupling length} $\lambda^{\star}$, that is much larger than the microscopic lattice spacing, i.e. $\lambda^{\star} \gg 1$ [we express length scales in units of the lattice spacing in $\mathcal{L}$]. In this case, we will proceed by first coarse-graining the microscopic model into an effective lattice model whose lattice spacing is of the order of $\lambda^{\star}$.

Let $\Wcoa$ denote the coarse-graining transformation of Ref. \cite{ER}, shown in Fig. \ref{fig:VandW}(b), which maps a lattice $\mathcal{L}$ with $N$ sites into an effective lattice $\mathcal{L}'$ with $N/2$ sites, and which uses \textit{disentanglers} $u$ in order to remove short-range entanglement, before applying \textit{isometries} $w$ to map two sites into one. [Those familiar with Refs. \cite{ER,MERA} may have noticed that $\Wcoa$ is a particular case of $\Wdec$ in which one of the two resulting lattices is in a product state]. We will use the scale parameter $s \equiv \log_2 (\lambda)$ to label the lattice $\mathcal{L}^{(s)}$ obtained after applying $s$ times the transformation $\Wcoa$ on $\mathcal{L}$. Then, in a system with \textit{decoupling scale} $s^{\star}\equiv \log_2 ( \lambda^{\star})$, we obtain a factorized description by: (i) applying a total of $s^{\star}$ coarse-graining transformations $W$ to the microscopic model, which result in a sequence of effective lattices $\{ \mathcal{L}^{(0)}, \mathcal{L}^{(1)},\cdots, \mathcal{L}^{(s^{\star})}\}$; (ii) applying the decoupling transformation $\Wdec$ on $\mathcal{L}^{(s^{\star})}$, which produces two decoupled models with lattices $\mathcal{L}^{(s^{\star}+1)}_A$ and $\mathcal{L}^{(s^{\star}+1)}_B$; and (iii) applying transformations $\Wcoa$ individually on each of the decoupled models to further coarse-grain them, as illustrated in Fig. \ref{fig:Y}.

Recall that the coarse-graining transformation $W$ alone produces the MERA \cite{ER,MERA}. In much the same way, if we collect all the two-site gates (disentanglers, isometries and swaps) contained in transformations $W$ and $V$ in Fig. \ref{fig:Y}, we obtain a new class of tensor network state, which branches into two independent tensor networks at scales $s$ larger than the decoupling scale $s^{\star}$ and which, accordingly, we refer to as the \textit{branching} MERA \cite{BranchingMERA}. This is a variational ansatz for states of the microscopic lattice $\mathcal{L}^{(0)}$, which inherits many of the computational advantages of the MERA, such as the efficient evaluation of local expectation values, e.g. energy and local order parameters, and of correlation functions.

\begin{figure}
  \begin{centering}
\includegraphics[width=8.5cm]{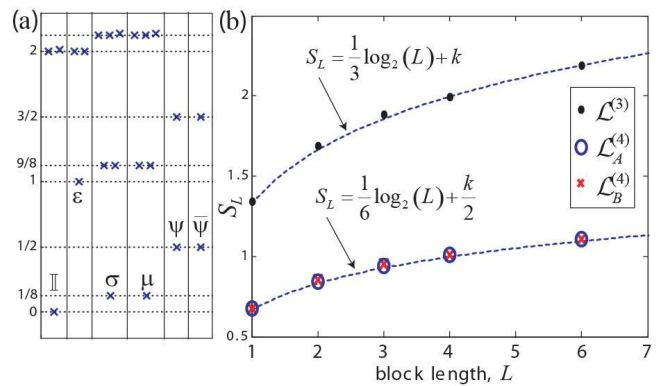}  \end{centering}
  \caption{
  (Colour online) (a) Scaling dimensions for scaling operators (extracted from a scale-invariant MERA \cite{CFT}) of each of the two decoupled spin chains. They are organized according to the conformal towers \cite{Yellow} of the critical quantum Ising model. (b) The entanglement entropy in each of the two decoupled chains (central charge $c=1/2$) is half of that in the system without decoupling (central charge $c=1$), see Eq. \ref{eq:addition}.
  }
  \label{fig:branches}
\end{figure}

To illustrate the performance of the approach, let us return to the critical quantum spin chain of Eq. \ref{eq:XX}. For $\mu \neq 0$, Hamiltonian $H^{\tiny\mbox{XX}}$ can no longer be decoupled analytically into two copies of Hamiltonian $H^{\tiny\mbox{Ising}}$, because the magnetic field $\frac{\mu}{2}\sum_r Z_r$ (or chemical potential in fermionic variables) turns into an interaction between the two sublattices. However, we managed to numerically factorize the ground state of the model. For instance, for $\mu = 2\cos(\pi/4)$ (or $1/8$ filling in the fermionic representation), an infinite lattice was first coarse-grained three times (i.e. $s^{\star} = 3$) with transformation $W$ and then decoupled into two chains with a transformation $V$ containing fermionic swaps. Then, assuming a factorized ground state (Eq. \ref{eq:fact_psi}) at that length scale, we used a scale-invariant MERA to find the ground state of each of the two reduced Hamiltonians (Eq. \ref{eq:fact_H2}). All variational parameters of the branching MERA were numerically optimized with simple variations of the standard, energy minimization algorithms of Refs. \cite{Algorithms,CFT}.

Specifically, using a local dimension $\chi=16$ ($\chi = 8$) for scales $s$ smaller than (respectively, larger than) the decoupling scale $s^{\star}=3$, the energy optimization took less than a day on a 2.66GHz workstation with 8GB of RAM. The resulting branching MERA approximated the exact ground state energy per site with an accuracy of $8\times 10^{-7}$ and duly captured the structure of large scale correlations in the system: from each branch we were able to extract \cite{CFT} accurate conformal data corresponding to the conformal field theory \cite{Yellow} that describes the critical Ising chain of Eq. \ref{eq:Ising}, see Fig. \ref{fig:branches}.

\begin{figure}[!tb]
\begin{center}
\includegraphics[width=8.5cm]{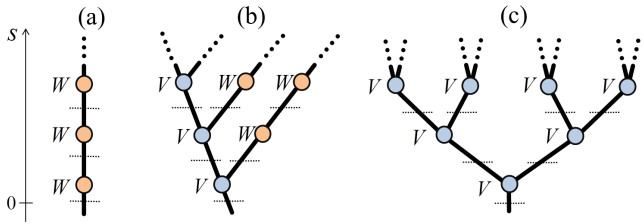}
\caption{
(color online) (a) Holographic tree without branching, representing the coarse-graining of a critical lattice model without separation of degrees of freedom. (b)-(c) Other possible holographic trees, with a regular pattern of branching. They describe fixed-points of a (generalized) RG flow where  separation of degrees of freedom occurs at all length scales.
} \label{fig:fixedpoint}
\end{center}
\end{figure}

\textit{Discussion.---} Two remarks are in order. First, in the example above decoupling was possible by using fermionic swaps (any other choice of particle statistics results in persistent interactions/correlations between the two sublattices), and applying $\Wdec$ at a proper scale $s^{\star}$. The particle statistics and decoupling scale $s^{\star}$ constitute important knowledge which we acquired by trial and error. Second, let $\rho_L$ denote the ground state reduced density matrix for a block of $L$ contiguous sites of $\mathcal{L}^{(0)}$. The entanglement entropy $S_L \equiv -\tr [\rho_L \log_2 (\rho_L)]$, which scales as $S_L \approx \frac{1}{3} \log(L)$ \cite{CriticalChain}, receives contributions from all length scales $\lambda \leq L$. Consider a block $L$ larger than the decoupling length $\lambda^{\star}$. Then it can be seen that the contribution to $S_L$ from a length scale $\lambda \in [\lambda^{\star}, L]$ is splitted evenly in contributions from the two decoupled chains (each of which asymptotically contributing $\frac{1}{6} \log (L)$, Fig. \ref{fig:branches}) which implies
\begin{equation}\label{eq:addition}
S_L \approx S^{(A)}_L + S^{(B)}_L, ~~~~~ L << \lambda^{\star}
\end{equation}
In other words, \textit{the (real-space) entanglement of a large block is roughly the sum of entanglements from each of the components into which the model decomposes}.

Generalizations to $D\geq 2$ dimensions are best discussed by introducing a $\textit{holographic tree}$,
Fig. \ref{fig:Y}(b), which captures the branching structure of a system under real space coarse-graining. In a general setting, we can imagine a holographic tree with a complex branching structure, corresponding to several instances of factorization of degrees of freedom occurring at various decoupling scales. This leads to an intriguing, multi-system generalization of the RG flow, with a variable number of independent many-body systems as a function of the scale parameter $s$. During the flow towards low energies, new systems may be added through branching.
In particular, a new notion of RG fixed point arises, corresponding to a self-similar holographic tree, see Fig. \ref{fig:fixedpoint}, which represents a many-body system with branching at all length scales.

A branching MERA in $D=2$ dimensions can describe the ground state of free fermions on a square lattice with a Fermi surface \cite{BranchingMERA}. The system seems to correspond to one of these novel, multi-branched RG fixed points. The entanglement entropy of a large block of $L\times L$ sites receives contributions from all relevant branches, generalizing Eq. \ref{eq:addition} in such a logarithmic violation of the boundary law, $S_{L} \approx L \log_2 (L)$, is reproduced. Accordingly, we expect the branching MERA to also provide an efficient description of Fermi liquids and spin-Bose metals.

G.E. is supported by the Sherman Fairchild Foundation. G.V. thanks the Australian Research Council Centre of Excellence for Engineered Quantum Systems.


\section{Appendix}
In this appendix we show that the XX quantum spin chain can be exactly decoupled into two (critical) quantum Ising spin chains by applying local disentanglers -- provided that fermionic degrees of freedom are used to separate the two systems. This derivation follows from the answer to exercise 12.1 in page 480 of Ref. \cite{Yellow}.

The XX quantum spin chain Hamiltonian without magnetic field is
\begin{equation}
	H = \sum_{r} X_r X_{r+1} + Y_r Y_{r+1}
\label{eq:Happendix}
\end{equation}
Let us introduce Majorana fermion operators $c_r$ and $d_r$, given by
\begin{equation}
	c_r \equiv \left(\prod_{l<r} Z_l\right) X_r, ~~~ d_r \equiv \left(\prod_{l<r} Z_l\right) Y_r.
\end{equation}
Notice that all $c_r$'s and $d_r$'s anticommute pairwise, except with themselves (since they fulfill $(c_r)^2 = (d_r)^2 = I$). In terms of these Majorana operators the Hamiltonian reads
\begin{equation}
	H = \sum_{r} i\left( c_r d_{r+1} - d_r c_{r+1}\right).
\end{equation}
Further, for each odd $r$, we apply a unitary gate $u$ on the pair of sites $(r,r+1)$, such that
\begin{eqnarray}
	 u ~c_r ~u^{\dagger} &=& c_{r+1}, ~~~
	 u ~d_r ~u^{\dagger} = c_r, ~~~ \nonumber\\
	 u ~c_{r+1} ~u^{\dagger} &=& d_r, ~~~
	 u ~d_{r+1} ~u^{\dagger} = -d_{r+1}. \label{eq:exact}
\end{eqnarray}
[Notice that $u$ preserves the anticommutation relations and it is therefore a canonical transformation of the fermionic variables.] Then the transformed Hamiltonian reads
\begin{equation}
  H = \sum_{r} h_{r,r+2}, ~~~~ h_{r,r+2} \equiv i\left( d_r c_{r+2} - c_r d_r \right),
\end{equation}
or $H = H_A + H_B$, where Hamiltonians $H_A$ and $H_B$,
\begin{equation}
	H_A \equiv \sum_{\tiny{\mbox{odd}} ~r} h_{r,r+2},~~~ H_B \equiv \sum_{\tiny{\mbox{even}} ~r} h_{r,r+2}
\end{equation}
commute with each other, since they are made of terms that are quadratic in Majorana fermion operators and they act on two different sets of sites.
Our decoupling transformation $\Wdec$ in Fig. \ref{fig:decoupling} will then consist of a row of disentanglers $u$ as in Eq. \ref{eq:exact}, followed by trivial disentanglers $v$ (i.e. $v=I$) since the system has already been decoupled.
To group together all sites of sublattice $\mathcal{L}_A$ to the left, and all sites of sublattice $\mathcal{L}_B$ to the right, we will use fermionic swap gates, Eq. \ref{eq:swap_f}. Notice that these swaps are trivial in fermionic variables, simply interchanging $(c_r,d_r)$ with $(c_{r+1},d_{r+1})$. However, they are non-trivial when expressed in spin variables as in Eq. \ref{eq:swap_f}.

\begin{figure}
  \begin{centering}
\includegraphics[width=6cm]{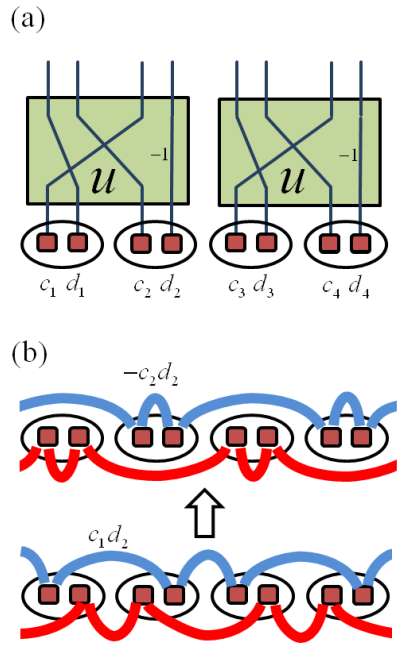}  \end{centering}
  \caption{
  (Colour online) (a) Representation of the exact disentangler $u$ of Eq. \ref{eq:exact} used to decouple Hamiltonian $H$ in Eq. \ref{eq:Happendix}. Each site, represented by an oval, contains two Majorana fermions, depicted as brown boxes. The disentangler $u$ permutes the Majorana modes and adds a minus sign to one of them. (b) Hamiltonian terms before (below) and after (above) the disentangler $u$ is applied on pairs of sites. Each Hamiltonian term, e.g. $c_1d_2$, is depicted with a line connecting two Majorana modes. Notice that after applying the disentangler, which e.g. maps $c_1d_2$ to $-c_2d_2$, the two sublattices are no longer connected by interactions.
}
  \label{fig:exact}
\end{figure}

Once the two lattices have been separated, we can define independent Jordan Wigner transformations on each lattice $\mathcal{L}^{(\alpha)}$, $\alpha = A,B$,
We can now define odd and even spin variables by
\begin{equation}
	c^{(\alpha)}_{r} = \left(\prod_{l<r} Z^{(\alpha)}_l\right) X^{(\alpha)}_r, ~~~ d^{(\alpha)}_{r} = \left(\prod_{l<r} Z^{(\alpha)}_l\right) Y^{(\alpha)}_r,
\end{equation}
in terms of which we obtain Hamiltonians
\begin{equation}
  H^{(\alpha)} = \sum_{r} \left( X^{(\alpha)}_rX^{(\alpha)}_{r+1} + Z^{(\alpha)}_r \right), ~~~\alpha = A,B,
\end{equation}
which correspond to two copies of the quantum Ising model with critical transverse magnetic field.

\end{document}